# On degenerate metrics, dark matter and unification

Trevor P. Searight[a]
*London, United Kingdom*

A five-dimensional theory of relativity is presented which suggests that gravitation and electromagnetism may be unified using a degenerate metric. There are four fields (in the four-dimensional sense): a tensor field, two vector fields and a scalar field, and they are unified with a combination of a gauge-like invariance and a reflection symmetry which means that both vector fields are photons. The gauge-like invariance implies that the fifth dimension is not directly observable; it also implies that charge is a constant of motion. The scalar field is analogous to the Brans-Dicke scalar field, and the theory tends towards the Einstein-Maxwell theory in the limit as the coupling constant tends to infinity. As there is some scope for fields to vary in the fifth dimension it is possible for the photons to have wave behaviour in the fifth dimension. The wave behaviour has two effects: it gives the photons mass, and it prevents them from interacting directly with normal matter. These massive photons still act as a source of gravity, however, and therefore they are candidates for dark matter. © *2017 Author(s)*.

## I. INTRODUCTION

The motivation for looking for a unified theory of gravitation and electromagnetism is well known: both Newton's law and Coulomb's law follow the inverse square rule, and charge is conserved as is energy-momentum. Nevertheless something is clearly different between the two phenomena since charge is quantized and is not carried by fields, in contrast to energy-momentum. So whilst one might hope to combine the two in a single framework, somehow differences must remain and these differences must be embedded within that framework. Perhaps the best known attempt at unification is Kaluza's theory[1] where electromagnetism is described by curvature in an extra spacelike dimension, however this theory has a flaw: it is supposed that the metric is independent of the fifth coordinate – a condition known as the cylinder condition – and the existence of a preferred coordinate breaks the unification. For while the Ricci scalar may contain the correct elements of both gravitation and electromagnetism, it is only unique as a scalar if physics is the same in all coordinate frames: if there is a preferred coordinate in five dimensions that allows new scalars to be constructed[2]. One cannot justify the use of the Ricci scalar in Kaluza's theory simply on the basis that it is what is used in general relativity. Various attempts have been made either to explain the cylinder condition or to improve upon Kaluza's theory – ranging from Klein's compactification[3] to projective relativity[4] to the uncompactified approach (see, for example, the review article Ref. 5) – however a definitive solution to the problem of unification has, seemingly, yet to be found.

Despite the shortcomings of the Kaluza-Klein theory, it is a logical idea to try to incorporate additional forces into general relativity by expanding beyond four dimensions. What is needed is a fully relativistic theory – one in which fields are allowed to vary in the fifth dimension – which separates the extra dimension from the other four while still unifying the gravitational and electromagnetic fields. The particular approach presented in this paper is to use a degenerate metric. The rationale for using a degenerate metric comes from the observation that the Hamiltonian for a charged particle in four dimensions

$$H = \frac{1}{2mc} g^{\mu\nu}(p_\mu - qA_\mu)(p_\nu - qA_\nu)$$

can be recast in five dimensions as

$$H = \frac{1}{2mc} \begin{pmatrix} p_\mu & \kappa q \end{pmatrix} \begin{pmatrix} g^{\mu\nu} & -\kappa^{-1} g^{\mu\rho} A_\rho \\ -\kappa^{-1} g^{\nu\sigma} A_\sigma & \kappa^{-2} g^{\rho\sigma} A_\rho A_\sigma \end{pmatrix} \begin{pmatrix} p_\nu \\ \kappa q \end{pmatrix}$$

where $\kappa$ is some constant: the determinant of the 5×5 matrix is zero. It is natural to wonder whether this is not just by chance but instead indicative of something more fundamental – a new theory of five dimensions in which the metric is degenerate. Just as Maxwell's equations were already Lorentz invariant before the discovery of special relativity, could the equations of motion – expressed this way in five dimensions with a degenerate metric – already be correct?

---

[a] Electronic mail: trevor.searight@btinternet.com

A degenerate metric implies a new type of dimension not seen in general relativity. In general relativity there are two types of dimension: spacelike, characterized by a value of +1 in the diagonal of the Minkowski metric (or –1, depending on convention), and timelike, characterized by a value of –1 in the diagonal (or +1, depending on convention). Coordinate invariance means that values of, say, +2, –2, etc. result in the same physics as +1, –1, etc. In summary, therefore, dimensions are characterized by solutions to the equation $x^2 = 1$: only the sign is important. With a degenerate metric a new type of dimension is possible which is characterized by a zero in the diagonal. It is still only the sign that is important, but now there are three possibilities: the solutions of $x^3 = x$. Recently Stoica has investigated degenerate metrics[6] and speculated that a degenerate metric might be used to explain why extra dimensions cannot be observed[7].

For a theory to be described as unified there should be a single, compelling approach where the only choices are the values of constants. Typically unification is achieved by requiring the laws of physics to be invariant under some transformation, and in Ref. 8 this author proposed a particular invariance which was an invariance of scale in the fifth dimension. By definition a degenerate metric has an eigenvector and there are two such eigenvectors – one each of the covariant metric and the contravariant metric – and the scale invariance was an invariance under a local scaling of the eigenvectors. However this led to difficulty recovering Coulomb's law: only source-free Maxwell's equations could be found. It was suggested that the electromagnetic field might act as its own source since the new field equations were non-linear, but the lack of suitable boundary conditions to justify the existence of these non-linear solutions over any trivial solution made the proposal unsatisfactory.

Instead it is proposed herein that the correct choice is to use a gauge-like invariance of the two eigenvectors to unify the field. There are a number of benefits to this solution. First, the invariance means that there is no preferred space-time direction at infinity; second, the invariance implies that charge is a constant of motion; third, it can be shown that matter density is independent of the fifth dimension, thus replacing the cylinder condition. A further symmetry is required to complete the unification and it is that both the vector fields represent photons; Coulomb's law is the result of these two fields acting together. As is common to higher-dimensional theories, there is an additional scalar field which is analogous to the Brans-Dicke scalar field, and the Einstein-Maxwell theory is recovered in the limit as the coupling constant tends to infinity.

Because fields are allowed to vary in the fifth dimension – albeit restricted by the field equations – there is the possibility of new phenomena arising from this additional degree of freedom. One such possibility is for photons to have wave behaviour in the fifth dimension. The wave behaviour results in two effects: it gives the photons mass, and it prevents them from interacting directly with normal matter (which has no dependency on the fifth dimension). There have already been various studies of the implications of a non-zero photon mass[9] which can be modelled using the Proca equation[10]. What these studies have in common is the premise that there is a single particle of light with a small, non-zero mass. The implication in the theory of degenerate metrics, however, is that there is a massive photon *in addition* to the massless one. Even though these massive photons do not interact directly with normal matter, they still act as a source of gravity, and therefore they are candidates for dark matter.

The problem of dark matter is sufficiently well known that it needs no detailed introduction. The rotation of distant galaxies cannot be explained by the visible matter they contain, leading to the conclusion that there must be some unseen matter within the galaxies exerting a gravitational force on the stars in the outer arms. Candidates for dark matter fall into three broad categories: very light particles called axions; very massive objects such as black holes or neutron stars; and weakly interacting massive particles ("WIMPs") – see, for example, the historical review article Ref. 11. Extra dimensions enter the subject in the form of "Kaluza-Klein dark matter"[12,13] where the size of the compact extra dimension gives the dark matter particles discrete masses. What distinguishes the theory of degenerate metrics is that the masses can take any value, since there is no fundamental length scale associated with the fifth dimension. That is not to say that the masses are not discrete, it is just that there is nothing in the geometry to make them so.

This paper is divided into eight sections, of which this introduction is the first. Section 2 begins by setting out the basics of degenerate metrics and defining the relevant tensor calculus. In section 3 the equations of motion are discussed, examining four different approaches: two using a Lagrangian, and two using a Hamiltonian. The field is unified in section 4. The first unification is based on a Weyl invariance, which is shown to be unsatisfactory; the second unification is based on a gauge invariance and it is this which is adopted for the remainder of the paper. The corresponding field equations are found in section 5. In section 6 the replacement for the cylinder condition is derived and there is a discussion about the exactness of the reflection symmetry. Section 7 solves the field equations for the massive photon and gives the accompanying energy-momentum tensor, from which the impact on the gravitational field can be inferred. There are some closing comments in section 8.

## II. DEFINITIONS

Denote the covariant metric by $\gamma_{ab}$ (replacing $g$ with $\gamma$ in five or more dimensions) and the contravariant metric by $\gamma^{ab}$. The covariant metric is degenerate and consequently obeys

$$\det \gamma_{ab} = 0 . \tag{1}$$

The contravariant metric can be partially defined by letting it satisfy

$$\gamma_{ab} \gamma^{bc} \gamma_{cd} = \gamma_{ad} . \tag{2}$$

If one additionally demands that the contravariant metric is degenerate then

$$\det \gamma^{ab} = 0 \tag{3}$$

and

$$\gamma^{ab} \gamma_{bc} \gamma^{cd} = \gamma^{ad} . \tag{4}$$

Now let $\varepsilon^{ab}$ be a matrix of eigenvectors of the covariant metric, such that

$$\gamma_{ab} \varepsilon^{bc} = 0 \tag{5}$$

and let $\varepsilon_{ab}$ be a matrix of eigenvectors of the contravariant metric

$$\gamma^{ab}\varepsilon_{bc} = 0 \tag{6}$$

such that the relation

$$\gamma^{ab}\gamma_{bc} + \varepsilon^{ab}\varepsilon_{bc} = \delta^a_c \tag{7}$$

holds. Both $\varepsilon^{ab}$ and $\varepsilon_{ab}$ are degenerate also, and they obey

$$\varepsilon^{ab}\varepsilon_{bc}\varepsilon^{cd} = \varepsilon^{ad} \tag{8}$$

and

$$\varepsilon_{ab}\varepsilon^{bc}\varepsilon_{cd} = \varepsilon_{ad}. \tag{9}$$

Now the purpose of the metric is to define how matter interacts with the field through the equations of motion. It is this that makes the metric "fundamental". But in addition to the metric (which is degenerate) it is possible to define a non-degenerate two-index tensor by

$$\tilde{\gamma}_{ab} = \gamma_{ab} + \lambda_0 \varepsilon_{ab} \tag{10}$$

where $\lambda_0$ is some non-zero dimensionless constant, and for some purposes it is more convenient to work with this quantity. In particular by using $\tilde{\gamma}_{ab}$ it means that the usual tensor calculus from general relativity can still be used. Thus for a contravariant vector $U^b$ define covariant derivative

$$\nabla_a U^b = \partial_a U^b + \Gamma^b_{ac} U^c \tag{11}$$

where $\Gamma^a_{bc}$ is the connection, for a covariant vector $V_b$ define covariant derivative

$$\nabla_a V_b = \partial_a V_b - \Gamma^c_{ab} V_c, \tag{12}$$

and let the product rule hold. The inverse of $\tilde{\gamma}_{ab}$ is $\tilde{\gamma}^{ab} = \gamma^{ab} + \lambda_0^{-1}\varepsilon^{ab}$. Define

$$\Gamma^a_{bc} = \tfrac{1}{2}\tilde{\gamma}^{ad}(\partial_b\tilde{\gamma}_{cd} + \partial_c\tilde{\gamma}_{bd} - \partial_d\tilde{\gamma}_{bc}) \tag{13}$$

(symmetric under $b \leftrightarrow c$) so that $\nabla_a\tilde{\gamma}_{bc} = 0$ and $\nabla_a\tilde{\gamma}^{bc} = 0$. As usual define Riemann tensor

$$R^a_{bcd} = \partial_c\Gamma^a_{bd} - \partial_b\Gamma^a_{cd} + \Gamma^e_{bd}\Gamma^a_{ce} - \Gamma^e_{cd}\Gamma^a_{be} \tag{14}$$

(antisymmetric under $b \leftrightarrow c$), Ricci tensor

$$R^a_c = R^a_{bcd}\tilde{\gamma}^{bd}, \tag{15}$$

and Ricci scalar

$$R = R^a_a. \tag{16}$$

Also define

$$\varphi = \sqrt{|\det(\tilde{\gamma}_{ab})|} \,. \tag{17}$$

Let us now work in a universe with five dimensions. Let indices $a$, $b$, $c$, ... run over 0, 1, 2, 3, 5 and let $\mu$, $\nu$, $\rho$, ... run over 0, 1, 2, 3 and denote the fifth dimension by $w$. In the absence of forces $\gamma_{ab} = \eta_{ab} \equiv$ diag $(-1, 1, 1, 1, 0)$ and $\varepsilon_{ab} =$ diag $(0, 0, 0, 0, 1)$. Because we work in five dimensions $\gamma^{ab}$ has a single eigenvector with eigenvalue zero, and so $\varepsilon_{ab}$ can be written as

$$\varepsilon_{ab} = \varepsilon_a \varepsilon_b \tag{18}$$

where

$$\gamma^{ab}\varepsilon_b = 0 \,. \tag{19}$$

Similarly write

$$\varepsilon^{ab} = \varepsilon^a \varepsilon^b \tag{20}$$

where

$$\gamma_{ab}\varepsilon^b = 0 \tag{21}$$

and

$$\varepsilon_a \varepsilon^a = 1 \,. \tag{22}$$

Define

$$E_{ab} = \partial_a \varepsilon_b - \partial_b \varepsilon_a \,, \tag{23}$$

an antisymmetric tensor;

$$E^{ab} = \gamma^{ac}(\partial_c \varepsilon^b) + \gamma^{bc}(\partial_c \varepsilon^a) - \varepsilon^c(\partial_c \gamma^{ab}) \,, \tag{24}$$

a symmetric tensor; and scalar $E = \gamma_{ab} E^{ab}$. It can be shown that

$$E = 2\varphi^{-1}\partial_a(\varphi \varepsilon^a) \,. \tag{25}$$

Define

$$\Phi_a = E_{ab}\varepsilon^b \,, \tag{26}$$

a vector, and corresponding scalar $\Phi = \gamma^{ab}\Phi_a\Phi_b$. Also define

$$h_b^a = \gamma^{ac}\gamma_{cb} \,, \tag{27}$$

a tensor. Note that because $h_b^a = \delta_b^a - \varepsilon^a \varepsilon_b$ there is no raising or lowering of indices with the notation in this paper.

## III. THE EQUATIONS OF MOTION

### A. The Lagrangian formulation

Let the path of a particle be parameterised by $x^a(s)$ and let $u^a = dx^a/ds$ be the 5-velocity. The first equations of motion to consider are those generated by the Lagrangian

$$L = \tfrac{1}{2} mc\gamma_{bc} u^b u^c + \kappa q \varepsilon_a u^a, \tag{28}$$

where $\kappa$ is some constant of nature (note that the $\kappa$ here replaces $\kappa^{-1}$ in Ref. 8). The 5-momentum is

$$p_a = mc\gamma_{ab} u^b + \kappa q \varepsilon_a \tag{29}$$

and it obeys

$$\varepsilon^a p_a = \kappa q, \tag{30}$$

so that the fifth component is proportional to charge. The equations of motion are obtained in the usual way from Lagrange's equation

$$\frac{d}{ds}\left(\frac{\partial L}{\partial u^a}\right) = \frac{\partial L}{\partial x^a}, \tag{31}$$

giving

$$\gamma_{ab} \frac{du^b}{ds} = \tfrac{1}{2}(\partial_a \gamma_{bc}) u^b u^c - (\partial_c \gamma_{ab}) u^b u^c + \frac{\kappa q}{mc} E_{ac} u^c. \tag{32}$$

Although equation (32) is a vector equation there are two scalar equations embedded in it which can be obtained by contracting with $u^a$ and $\varepsilon^a$. Firstly, the value of the Lagrangian $L(s)$ can be shown to be constant in the usual way, so that

$$\gamma_{ab} u^a u^b = \text{constant}; \tag{33}$$

the constant can be set to $-1$ with a suitable scaling of $s$. Secondly, contracting equation (32) with $\varepsilon^a$ gives

$$\tfrac{1}{2}\varepsilon^a(\partial_a \gamma_{bc}) u^b u^c - \varepsilon^a(\partial_c \gamma_{ab}) u^b u^c + \frac{\kappa q}{mc} \varepsilon^a E_{ac} u^c = 0, \tag{34}$$

which can be written in fully covariant form as

$$\tfrac{1}{2} E^{ab} \gamma_{ac} \gamma_{bd} u^c u^d - \frac{\kappa q}{mc} \Phi_c u^c = 0. \tag{35}$$

As this equation has no term in the derivative $du^b/ds$ it would seem to represent a constraint – either on the particle or on the field. One might have expected that equation (35) would determine $\varepsilon_a u^a$, however it in fact says nothing at all about $\varepsilon_a u^a$ since $u^a$ may be replaced by $u^a - \varepsilon^a(\varepsilon_c u^c)$ without altering the equation. Because a constraint on the four-velocity $h_b^a u^b$ would be unphysical, it suggests that equation (35) is solved by the field in some way. Equation (35) may be rewritten as

$$E^{ab} p_a p_b = 0 \tag{36}$$

since $\Phi_c = -\varepsilon_a E^{ab} \gamma_{bc}$. The simplest solution would therefore be

$$E^{ab} = 0 \tag{37}$$

in which case $\varepsilon^a$ would be a Killing vector, as equation (37) is equivalent to

$$\gamma^{ac} \nabla_c \varepsilon^b + \gamma^{bc} \nabla_c \varepsilon^a = 0. \tag{38}$$

Whether $\varepsilon^a$ is a Killing vector or not, equation (32) represents four equations for the five unknowns $u^a$, since equation (35) is a constraint on the field and not the particle. Equation (32) may be written in covariant form as

$$h_b^a \frac{du^b}{ds} + h_b^a \Gamma_{cd}^b u^c u^d = \left( \frac{\kappa q}{mc} - \lambda_0 \varepsilon_d u^d \right) \gamma^{ab} E_{bc} u^c. \tag{39}$$

Maxwell's equations are easily recovered by working in (4+1)-dimensional notation. Write

$$\gamma^{ab} = \begin{pmatrix} g^{\mu\nu} & -g^{\mu\rho}\varepsilon_\rho \varepsilon_5^{-1} \\ -g^{\nu\sigma}\varepsilon_\sigma \varepsilon_5^{-1} & (g^{\rho\sigma}\varepsilon_\rho \varepsilon_\sigma)\varepsilon_5^{-2} \end{pmatrix}$$

$$\varepsilon_a = (\varepsilon_\mu, \quad \varepsilon_5) \tag{40}$$

$$\varepsilon^a = (\varepsilon^\mu, \quad \varepsilon_5^{-1} - \varepsilon_\rho \varepsilon^\rho \varepsilon_5^{-1})$$

$$\gamma_{ab} = \begin{pmatrix} g_{\mu\nu} - g_{\mu\rho}\varepsilon^\rho \varepsilon_\nu - g_{\nu\rho}\varepsilon^\rho \varepsilon_\mu + (g_{\rho\sigma}\varepsilon^\rho \varepsilon^\sigma)\varepsilon_\mu \varepsilon_\nu & -g_{\mu\rho}\varepsilon^\rho \varepsilon_5 + (g_{\rho\sigma}\varepsilon^\rho \varepsilon^\sigma)\varepsilon_\mu \varepsilon_5 \\ -g_{\nu\sigma}\varepsilon^\sigma \varepsilon_5 + (g_{\rho\sigma}\varepsilon^\rho \varepsilon^\sigma)\varepsilon_\nu \varepsilon_5 & (g_{\rho\sigma}\varepsilon^\rho \varepsilon^\sigma)(\varepsilon_5)^2 \end{pmatrix}.$$

For simplicity consider the case where $g^{\mu\nu} = \eta^{\mu\nu}$ and $\varepsilon_5 = 1$. Then ignoring second order terms and terms in $\partial_5$, equation (39) implies

$$\frac{du^\mu}{ds} = \frac{\kappa q}{mc} g^{\mu\nu}(\partial_\nu \varepsilon_\rho - \partial_\rho \varepsilon_\nu)u^\rho - g^{\mu\nu}(\partial_\nu(g_{\rho\sigma}\varepsilon^\sigma) - \partial_\rho(g_{\nu\sigma}\varepsilon^\sigma))u^\rho u^5 \tag{41}$$

which is the familiar form with electromagnetic field $\varepsilon_\nu$ coupled to charge $\kappa q$ and the field $-g_{\nu\sigma}\varepsilon^\sigma$ coupled to $mcu^5$.

The obvious benefit of this variant is that charge is constant, however there are two problems. First, there are only four equations for the five unknowns $u^a$: there is no equation of motion for $\varepsilon_a u^a$. Second, if $\varepsilon_a u^a$ (i.e. $u^5$) is non-zero, then from equation (41) the quantity $mcu^5$ behaves like an extra, unwanted charge. Furthermore, if $\varepsilon_a u^a$ is anything like $\kappa q/mc$, the quantity $q^2/m$ will act as a source of some scalar field via the term $\kappa q \varepsilon_a u^a$ in the Lagrangian (28), and this contradicts experience. It is well known that the force between charges is considerably greater than the force between masses; for a proton

$$\frac{q_p}{m_p}\frac{1}{\sqrt{4\pi\varepsilon_0 G}}=1.1\times 10^{18}.$$

If there were any field with $q^2/m$ as its source it would dominate over all others. Conclude that equations of motion derived from the Lagrangian (28) may be viable, but only if $\varepsilon_a u^a$ is zero (by a means as yet unspecified) and provided that the constraint (36) is understood.

An alternative to setting $\varepsilon_a u^a$ equal to zero is to set $q$ to zero, i.e. to use the Lagrangian

$$L = \tfrac{1}{2}mc\gamma_{bc}u^b u^c. \tag{42}$$

The equations of motion in covariant form are equation (39) without the term in $q$, and equation (36). In the (4+1)-dimensional notation

$$\frac{du^\mu}{ds} = -g^{\mu\nu}(\partial_\nu(g_{\rho\sigma}\varepsilon^\sigma) - \partial_\rho(g_{\nu\sigma}\varepsilon^\sigma))u^\rho u^5 \tag{43}$$

(assuming weak fields, etc.) suggesting that $\varepsilon_a u^a$ is $\kappa q/mc$. Now $p_5$ is zero, so there is no unwanted source of $q^2/m$, however there are still only four equations for the five unknowns $u^a$, and there is still the constraint on the field given by equation (36) to understand.

## B. The Hamiltonian formulation

In addition to a Lagrangian one can consider the Hamiltonian formulation of the equations of motion. Typically for every Lagrangian there is a corresponding Hamiltonian

$$H = u^a p_a - L \tag{44}$$

which can be used to generate the same equations of motion using Hamilton's equations

$$u^a = \frac{\partial H}{\partial p_a} \tag{45}$$

and

$$\frac{dp_a}{ds} = -\frac{\partial H}{\partial x^a}. \tag{46}$$

However, where the metric is degenerate the Hamiltonian formulation leads to different results from the Lagrangian formulation.

The Hamiltonian that is obtained by applying equation (44) to the Lagrangian (28) is

$$H = \frac{1}{2mc}\gamma^{ab}p_a p_b, \tag{47}$$

whence

$$u^a = \frac{1}{mc}\gamma^{ab}p_b \tag{48}$$

and

$$\frac{dp_a}{ds} = -\frac{1}{2mc}(\partial_a \gamma^{bc})p_b p_c. \tag{49}$$

These are the equations of motion that were posited in Ref. 8. They may be written in covariant form as

$$\frac{d}{ds}(\varepsilon^a p_a) = \frac{1}{2mc} E^{ab} p_a p_b \tag{50}$$

and

$$h_b^a \frac{du^b}{ds} + h_b^a \Gamma_{cd}^b u^c u^d = \frac{1}{mc}(\varepsilon^d p_d)\gamma^{ab} E_{bc} u^c. \tag{51}$$

Equation (51) is the same as equation (39) but with $\kappa q$ replaced by $\varepsilon^d p_d$ and $\varepsilon_d u^d$ removed, however there is no longer a constraint of the type given by equation (36). The only constraint is on the particle, for contracting equation (48) with $\varepsilon_a$ gives

$$\varepsilon_a u^a = 0. \tag{52}$$

Thus particles remain at fixed points in the fifth dimension, though they may not all be at the same point in the fifth dimension as each other: if fields are allowed to vary with $w$, that raises the question of why the variation in field strength in the fifth dimension is not observed in the real world. Using the (4+1)-dimensional notation equation (51) implies that

$$\frac{du^\mu}{ds} = \frac{1}{mc} p_5 g^{\mu\nu}(\partial_\nu \varepsilon_\rho - \partial_\rho \varepsilon_\nu)u^\rho \tag{53}$$

(assuming weak fields, etc.) which is the familiar form with electromagnetic field $\varepsilon_\nu$ coupled to $p_5$. As usual the value of the Hamiltonian $H(s)$ is a constant of motion, i.e.

$$\gamma^{ab} p_a p_b = \text{constant}; \tag{54}$$

the constant can be set to $-(mc)^2$ with a suitable scaling of $s$.

The benefit of using the Hamiltonian is that it gives a complete description of the motion, since there are five equations for $u^a$ and five equations for $p_a$, in contrast with the Lagrangian (28) which results in only four equations for $u^a$. However the nature of charge is unclear. If $\varepsilon^a p_a$ is constant then equation (50) implies that $E^{ab} p_a p_b = 0$ which suggests the same constraint on the field as was discussed in section 3A, i.e. $E^{ab} = 0$; if $\varepsilon^a p_a$ is not constant then, to be consistent with experiment, any variation in it must be negligible, and therefore a mechanism needs to be established such that $E^{ab} p_a p_b$ remains small. This issue will be addressed in later sections.

The last possibility to consider is the Hamiltonian

$$H = \frac{1}{2mc}\gamma^{ab} p_a p_b + \frac{\kappa q}{mc}\varepsilon^a p_a. \tag{55}$$

The equations of motion are

$$u^a = \frac{1}{mc}\gamma^{ab}p_b + \frac{\kappa q}{mc}\varepsilon^a \tag{56}$$

and

$$\frac{dp_a}{ds} = -\frac{1}{2mc}(\partial_a\gamma^{bc})p_b p_c - \frac{\kappa q}{mc}(\partial_a\varepsilon^b)p_b. \tag{57}$$

In covariant form these are equation (50) and

$$h_b^a \frac{du^b}{ds} + h_b^a \Gamma_{cd}^b u^c u^d = \left(\frac{\varepsilon^d p_d}{mc} - \lambda_0 \frac{\kappa q}{mc}\right)\gamma^{ab}E_{bc}u^c. \tag{58}$$

In the (4+1)-dimensional notation

$$\frac{du^\mu}{ds} = \frac{1}{mc}p_5 g^{\mu\nu}(\partial_\nu\varepsilon_\rho - \partial_\rho\varepsilon_\nu)u^\rho - \frac{\kappa q}{mc}g^{\mu\nu}(\partial_\nu(g_{\rho\sigma}\varepsilon^\sigma) - \partial_\rho(g_{\nu\sigma}\varepsilon^\sigma))u^\rho, \tag{59}$$

(assuming weak fields, etc.) implying that $p_5$ must be zero (but with no mechanism to impose this).

## IV. UNIFICATION

### A. Approach

There are four fields (in the four-dimensional sense) in the five-dimensional theory of degenerate metrics: a tensor field, two vector fields and a scalar field (the tensor and one vector from the covariant metric, one further vector from the contravariant metric, and the scalar from $\varphi$). The field equations are obtained in the usual way by varying a Lagrangian density $\mathcal{L}$

$$\delta\int \mathcal{L}\varphi dV = 0 \tag{60}$$

where $dV$ is the volume element. Unification, for this author, means that there is a single, valid Lagrangian density. The four-dimensional Einstein-Maxwell theory cannot be considered unified because there is no reason why the Ricci scalar should simply be combined with $g^{\mu\rho}g^{\nu\sigma}F_{\mu\nu}F_{\rho\sigma}$ in a linear sum. There is no *conceptual* reason why it should not be $(g^{\mu\rho}g^{\nu\sigma}F_{\mu\nu}F_{\rho\sigma})^2$ or some other function of $g^{\mu\rho}g^{\nu\sigma}F_{\mu\nu}F_{\rho\sigma}$ – even if, in practical terms, this might not give the desired result.

There are seven scalars involving the second derivative of the field in the theory of degenerate metrics: $R$, $\Phi$, $\gamma^{ab}\gamma^{cd}E_{ac}E_{bd}$, $\gamma_{ab}\gamma_{cd}E^{ac}E^{bd}$, $E^2$, $\varphi^{-1}\partial_a(\varphi\gamma^{ab}\Phi_b)$ and $\varphi^{-1}\partial_a(\varphi\varepsilon^a E)$, with the two divergences being of no interest with regard to building the Lagrangian density. There are two distinct ways to combine these scalars, which are described in the following sections.

### B. Weyl invariance

The first invariance to consider is invariance under the transformation

$$\gamma_{ab} \to \lambda\gamma_{ab}$$

$$\varepsilon_a \to \lambda^{-1}\varepsilon_a$$

$$\tag{61}$$

$$\gamma^{ab} \to \lambda^{-1}\gamma^{ab}$$

$$\varepsilon^a \to \lambda\varepsilon^a$$

which is an extension of the Weyl transformation[14]. Under (61) one finds that

$$\varphi \to \lambda\varphi \qquad (62)$$

using the relation $\varphi^{-1}(\partial_a\varphi) = \tfrac{1}{2}\gamma^{bc}(\partial_a\gamma_{bc}) + \varepsilon^b(\partial_a\varepsilon_b)$. It follows that Lagrangian densities transform as $\mathcal{L} \to \lambda^{-1}\mathcal{L}$ if the laws of physics are to be invariant under the Weyl transformation. This yields an immediate result, as the cosmological constant $\Lambda$ must consequently be zero.

Under the Weyl transformation

$$R \to \lambda^{-1}R + 3\lambda^{-2}\gamma^{ab}(\partial_a\lambda)\Phi_b - \tfrac{3}{2}\lambda^{-3}\gamma^{ab}(\partial_a\lambda)(\partial_b\lambda) - \lambda^{-1}\varphi^{-1}\partial_a(\varphi\gamma^{ab}\lambda^{-1}(\partial_b\lambda))$$
$$+ \tfrac{1}{4}\lambda_0(\lambda^{-1} - \lambda^{-4})\gamma^{ab}\gamma^{cd}E_{ac}E_{bd}$$
$$+ \tfrac{1}{4}\lambda_0^{-1}(\lambda^{-1} - \lambda^2)(\gamma_{ab}\gamma_{cd}E^{ac}E^{bd} - E^2) + \tfrac{3}{2}\lambda_0^{-1}\lambda\varepsilon^a(\partial_a\lambda)E + 3\lambda_0^{-1}(\varepsilon^a(\partial_a\lambda))^2$$
$$+ \lambda_0^{-1}\lambda^{-1}\varphi^{-1}\partial_a(\varphi\varepsilon^a E) - \lambda_0^{-1}\lambda^{-1}\varphi^{-1}\partial_a(\varphi\varepsilon^a\lambda^3 E) - 4\lambda_0^{-1}\lambda^{-1}\varphi^{-1}\partial_a(\varphi\varepsilon^a\lambda^2\varepsilon^b(\partial_b\lambda)) ;$$

$$\gamma^{ab}\gamma^{cd}E_{ac}E_{bd} \to \lambda^{-4}\gamma^{ab}\gamma^{cd}E_{ac}E_{bd} ;$$

$$\gamma_{ab}\gamma_{cd}E^{ac}E^{bd} \to \lambda^2\gamma_{ab}\gamma_{cd}E^{ac}E^{bd} + 2\lambda E\varepsilon^a(\partial_a\lambda) + 4(\varepsilon^a(\partial_a\lambda))^2 ;$$

$$E \to \lambda E + 4\varepsilon^a(\partial_a\lambda) ;$$

and

$$\Phi \to \lambda^{-1}\Phi - 2\lambda^{-2}\gamma^{ab}(\partial_a\lambda)\Phi_b + \lambda^{-3}\gamma^{ab}(\partial_a\lambda)(\partial_b\lambda)$$

whence

$$\mathcal{L} = \frac{c^3}{16\pi G}(R + \tfrac{3}{2}\Phi + \tfrac{1}{4}\lambda_0\gamma^{ab}\gamma^{cd}E_{ac}E_{bd} + \tfrac{1}{4}\lambda_0^{-1}\gamma_{ab}\gamma_{cd}E^{ac}E^{bd} - \tfrac{1}{4}\lambda_0^{-1}E^2) . \qquad (63)$$

The choices in (61) were not accidental, for given that $\gamma_{ab} \to \lambda\gamma_{ab}$ it is necessary that $\varepsilon_a \to \lambda^{-1}\varepsilon_a$ (as opposed to any other power of $\lambda$). For in order to recover general relativity the Lagrangian must be comprised of terms involving two partial derivatives $\partial.\partial.$ contracted with a number of covariant metrics $\gamma_{..}$ and contravariant metrics $\gamma^{..}$ such that the total number of raised indices equals the total number of lowered indices. In any one term, therefore, there will be one more contravariant metric than covariant metric, and so if $\gamma^{ab} \to \lambda^{-1}\gamma^{ab}$ then $\mathcal{L} \to \lambda^{-1}\mathcal{L}$. It follows that $\varphi \to \lambda\varphi$ and therefore that $\varepsilon_a \to \lambda^{-1}\varepsilon_a$. Note that the Lagrangian (63) is not unified in the strictest sense, because it is possible to add, for example, the term $\sqrt{(\gamma^{ab}\gamma^{cd}E_{ac}E_{bd})(\gamma_{ab}\gamma_{cd}E^{ac}E^{bd} - \tfrac{1}{4}E^2)}$ while keeping Weyl invariance – though the resulting Lagrangian would undoubtedly fail a practicality test.

For the equations of motion the obvious choice is to try the Weyl-invariant Lagrangian

$$L = k(\varepsilon_a u^a)(\gamma_{bc}u^b u^c) \qquad (64)$$

where $k$ is a constant (the only choice if one is to avoid massless particles). However one finds with this Lagrangian that Coulomb's law cannot be recovered. For consider a static, weak field

$$\gamma_{ab} = \begin{pmatrix} -1 & 0 & \chi \\ 0 & \delta_{ij} & 0 \\ \chi & 0 & -\chi^2 \end{pmatrix} \qquad \gamma^{ab} = \begin{pmatrix} -1 & 0 & \psi \\ 0 & \delta^{ij} & 0 \\ \psi & 0 & -\psi^2 \end{pmatrix} \qquad \varphi = 1.$$

Then

$$L = k(\psi u^0 + u^5)(-u^0 u^0 + u^i u^i + 2\chi u^0 u^5). \tag{65}$$

The dominant term in the Lagrangian density (63) is one in $(\partial_i \psi)(\partial_i \chi)$. It follows that the source of the field $\psi$ is obtained by varying equation (64) with respect to $\chi$, while the source of $\chi$ is obtained by varying with respect to $\psi$. Thus the source of $\psi$ is

$$2k u^0 (u^5)^2 \rho,$$

where $\rho$ is the particle density, while the source of $\chi$ is

$$-k(u^0)^3 \rho.$$

By contrast the quantities which these fields act upon are reversed: from equation (64) $\psi$ acts upon $-k(u^0)^3 \rho$ while $\chi$ acts upon $2k u^0 (u^5)^2 \rho$. There is no arrangement of $k$, $u^0$ and $u^5$ for which the combined effect of these two fields equates to Coulomb's law. For example, if $u^0 = 1$, $u^5 = \kappa q / mc$ and $k = \frac{1}{2}(mc)^2 / \kappa q$ then the force between two particles from the two vector fields is proportional to

$$m_1 m_2 \left( \frac{m_1 q_1}{m_2 q_2} + \frac{m_2 q_2}{m_1 q_1} \right).$$

Therefore conclude that this modified form of Weyl invariance is not satisfactory.

## C. 'Eigengauge' invariance

Another invariance to consider is gauge invariance. It makes good physical sense for the field to be invariant under a gauge transformation as such an invariance ensures that there is no preferred direction in space-time at infinity. In four dimensions coordinate invariance means that at infinity the metric $g_{\mu\nu}$ can be diagonalised to the Minkowski metric, and in five dimensions the same applies to the quantity $\tilde{\gamma}_{ab}$, however that still leaves the vector $\varepsilon_a$ remaining which may take non-trivial values in its first four components at infinity. The gauge invariance can be used to set those values to zero.

Therefore consider the following transformation, to be called the *eigengauge transformation*:

$$\gamma_{ab} \to \gamma_{ab} - \lambda_0 (\partial_a \alpha) \varepsilon_b - \lambda_0 (\partial_b \alpha) \varepsilon_a - \lambda_0 (\partial_a \alpha)(\partial_b \alpha)$$

$$\varepsilon_a \to \varepsilon_a + (\partial_a \alpha)$$

$$\varepsilon^a \to \lambda \varepsilon^a + \lambda_0 \gamma^{ab}(\partial_b \alpha) \tag{66}$$

$$\gamma^{ab} \to \gamma^{ab} - \lambda \gamma^{ac}(\partial_c \alpha)\varepsilon^b - \lambda \gamma^{bc}(\partial_c \alpha)\varepsilon^a - \lambda_0 \gamma^{ac}(\partial_c \alpha)\gamma^{bd}(\partial_d \alpha) + \gamma^{cd}(\partial_c \alpha)(\partial_d \alpha)\varepsilon^a \varepsilon^b$$

where $\lambda = 1 + \varepsilon^c(\partial_c \alpha)$ and

$$\lambda^2 + \lambda_0 \gamma^{ab}(\partial_a \alpha)(\partial_b \alpha) = 1 \tag{67}$$

so that the relation $\varepsilon_a \varepsilon^a = 1$ is preserved. Equation (67) represents a constraint upon the function $\alpha$, as it now cannot be any function. This is permissible since, to remove the preferred direction at infinity, there are only four vector components to be set to zero, and not five, and so there is a degree of freedom remaining which means there is room for a constraint. Let invariance under the eigengauge transformation be called *eigengauge invariance*.

By design $\tilde{\gamma}_{ab}$ is eigengauge invariant and so $\varphi$ and the Christoffel symbols $\Gamma^a_{bc}$ are also eigengauge invariant. Thus under the eigengauge transformation

$$R \to R; \tag{68}$$

$$\gamma^{ab}\Phi_a\Phi_b \to \gamma^{ab}\Phi_a\Phi_b - 2\lambda_0 \gamma^{ab}\gamma^{cd}(\partial_a\alpha)E_{bc}\Phi_d + O(\alpha^2); \tag{69}$$

$$\gamma^{ab}\gamma^{cd}E_{ac}E_{bd} \to \gamma^{ab}\gamma^{cd}E_{ac}E_{bd} + 4\gamma^{ab}\gamma^{cd}(\partial_a\alpha)E_{bc}\Phi_d + O(\alpha^2); \tag{70}$$

$$\gamma_{ab}\gamma_{cd}E^{ac}E^{bd} \to \gamma_{ab}\gamma_{cd}E^{ac}E^{bd} + 4\lambda_0 E^{ab}\nabla_a(\partial_b\alpha)$$
$$+ 4\lambda_0^2 \gamma^{ab}\gamma^{cd}(\partial_a\alpha)E_{bc}\Phi_d + O(\alpha^2); \tag{71}$$

$$E \to E + 2\lambda_0 \varphi^{-1}\partial_a(\varphi\gamma^{ab}(\partial_b\alpha)) + O(\alpha^2); \tag{72}$$

$$\varphi^{-1}\partial_a(\varphi\gamma^{ab}\Phi_b) \to \varphi^{-1}\partial_a(\varphi\gamma^{ab}\Phi_b) - \tfrac{1}{2}E\gamma^{ab}(\partial_a\alpha)\Phi_b + \tfrac{1}{2}\lambda_0\Phi_a\gamma^{ab}E_{bc}\gamma^{ce}(\partial_e\alpha)$$
$$- \varepsilon^a(\nabla_a\Phi_b)\gamma^{bc}(\partial_c\alpha) + \tfrac{1}{2}E^{ab}(\partial_a\alpha)\Phi_b + \lambda_0\nabla_a(\gamma^{ab}E_{bc}\gamma^{cd}(\partial_d\alpha)) + O(\alpha^2); \tag{73}$$

and

$$\varphi^{-1}\partial_a(\varphi\varepsilon^a E) \to \varphi^{-1}\partial_a(\varphi\varepsilon^a E) + 2\lambda_0 E\varphi^{-1}\partial_a(\varphi\gamma^{ab}(\partial_b\alpha))$$
$$+ 2\lambda_0 \gamma^{ab}(\partial_a\alpha)(\partial_b E) - 2\lambda_0 \nabla_a(E^{ab}(\partial_b\alpha)) + O(\alpha^2). \tag{74}$$

The following two scalars are eigengauge invariant: $R$ and

$$\gamma^{ab}\Phi_a\Phi_b + \tfrac{1}{2}\lambda_0\gamma^{ab}\gamma^{cd}E_{ac}E_{bd},$$

with the latter easily seen to be invariant since it is equal to $\tilde{\gamma}^{ab}\tilde{\gamma}^{cd}E_{ac}E_{bd}$ and $E_{ab}$ is invariant. Therefore the most general Lagrangian which is consistent with the eigengauge transformation is

$$\mathcal{L} = \frac{c^3}{16\pi G_0}(R - 2\Lambda - \omega\gamma^{ab}\Phi_a\Phi_b - \tfrac{1}{2}\omega\lambda_0\gamma^{ab}\gamma^{cd}E_{ac}E_{bd}) \tag{75}$$

where $\omega$ is a dimensionless constant, $\Lambda$ is the cosmological constant, and $G_0$ is a constant with the same dimensions as Newton's constant of gravitation $G$.

Requiring the field to be invariant under the eigengauge transformation achieves a partial unification: the four fields (tensor, scalar and two vectors) have been condensed into just two scalar quantities (ignoring the cosmological constant). As for matter, neither the Lagrangians (28) and (42) nor the Hamiltonians (47) and (55) are eigengauge invariant. For example, the Hamiltonian (47) transforms as

$$H \to H - \frac{1}{mc}(\varepsilon^a p_a)(\partial_b \alpha)\gamma^{bc} p_c + O(\alpha^2)$$
$$= H - (\varepsilon^a p_a)(\partial_b \alpha)u^b + O(\alpha^2). \qquad (76)$$

Thus it cannot be said that physics is invariant under the eigengauge transformation; only the field equations *in vacuo* are invariant. But because the field is always eigengauge invariant, a further equation of motion can be obtained by applying the eigengauge transformation to either the Lagrangian or the Hamiltonian for matter and then making use of the variational principle. Rewriting equation (76) as

$$H \to H - \frac{d}{ds}(\alpha \varepsilon^a p_a) + \alpha \frac{d}{ds}(\varepsilon^a p_a) + O(\alpha^2) \qquad (77)$$

we can deduce that $\varepsilon^a p_a$ is constant; equation (50) then gives $E^{ab} p_a p_b = 0$, i.e. equation (36). Similar results apply to the Lagrangians (28) and (42) and the Hamiltonian (55). For example, under the eigengauge transformation the Lagrangian (42) transforms as

$$L \to L - \lambda_0 mc(\varepsilon_a u^a)(\partial_b \alpha)u^b + O(\alpha^2) \qquad (78)$$

implying that $\varepsilon_a u^a$ is constant. Thus there is no longer a "missing" equation of motion for either Lagrangian (28) or (42): there are now five equations for the five unknowns $u^a$.

With eigengauge invariance the four different approaches to the equations of motion – the Lagrangians (28) and (42) and the Hamiltonians (47) and (55) – all take the same general form, so that they may be summarized as follows:

$$h^a_b u^b = \frac{1}{mc} \gamma^{ab} p_b; \qquad (79)$$

$$h^a_b \frac{du^b}{ds} + h^a_b \Gamma^b_{cd} u^c u^d = \left(\frac{\varepsilon^d p_d}{mc} - \lambda_0 \varepsilon_d u^d\right) \gamma^{ab} E_{bc} u^c; \qquad (80)$$

$$E^{ab} p_a p_b = 0;$$

$$\varepsilon^a p_a = \text{constant}; \qquad (81)$$

$$\varepsilon_a u^a = \text{constant}; \qquad (82)$$

and

$$(\varepsilon_a u^a)(\varepsilon^b p_b) = 0 \qquad (83)$$

in order to eliminate the source of the scalar field. The Lagrangian (28) is equivalent to the Hamiltonian (47), while the Lagrangian (42) is equivalent to the Hamiltonian (55).

Returning to the notation introduced at the end of section 4A and considering purely electromagnetic terms, equation (75) may be written

$$\mathcal{L} = \frac{c^3}{16\pi G_0}\left(\tfrac{1}{2}\lambda_0(2\omega+1)(\partial_i \psi)(\partial_i \psi) + (\partial_i \psi)(\partial_i \chi) + \frac{1}{2\lambda_0}(\partial_i \chi)(\partial_i \chi)\right)$$

plus higher order terms. This expression may be rewritten as

$$\mathcal{L} = \frac{c^3}{16\pi G_0} \tfrac{1}{2}\lambda_0 \begin{pmatrix} \partial_i \psi & \partial_i \psi + \lambda_0^{-1}\partial_i \chi \end{pmatrix} \begin{pmatrix} 2\omega & 0 \\ 0 & 1 \end{pmatrix} \begin{pmatrix} \partial_i \psi \\ \partial_i \psi + \lambda_0^{-1}\partial_i \chi \end{pmatrix}. \tag{84}$$

Thus in the case $\lambda_0 > 0$, $\omega > 0$ the Lagrangian (75) combines two positive energy photons; in the case $\lambda_0 > 0$, $\omega < 0$ the Lagrangian combines a positive energy photon with a negative energy photon; in the case $\lambda_0 < 0$, $\omega < 0$ the Lagrangian also combines a positive energy photon with a negative energy photon (but with the two reversed from the case $\lambda_0 > 0$, $\omega < 0$); while in the case $\lambda_0 < 0$, $\omega > 0$ the Lagrangian combines two negative energy photons.

As regards a full unification of the field, the Lagrangian (75) is the linear sum of two eigengauge invariant scalars (ignoring the cosmological constant). In order to be able to say the field is unified this sum must be justified, otherwise this Lagrangian is no more unified that the four-dimensional Einstein-Maxwell one. However the five-dimensional Lagrangian does have a unique property. As it has just been shown, the Lagrangian (75) describes two photons: $\varepsilon_a$ and $\varepsilon^a$, and the two vector fields can only both be photons with a linear sum. Furthermore, with a linear sum the two vectors fields are (to first order) interchangeable, so that it does not matter which of $\varepsilon_a u^a$ and $\varepsilon^a p_a$ is non-zero, since both generate Maxwell's equations, provided that $\lambda_0 > 0$ and $\omega > 0$. In the case of Lagrangian (28) and Hamiltonian (47) charge is represented by the fifth component of the 5-momentum, while in the case of Lagrangian (42) and Hamiltonian (55) it is represented by the fifth component of the 5-velocity. In each case only one of the two vector fields $\varepsilon_a$ and $\varepsilon^a$ interacts with matter: one might term the field that interacts with matter "radiant" light and the one that does not interact with matter "dark" light. In the case of Lagrangian (28) and Hamiltonian (47) it is $\varepsilon_a$ which interacts with matter, while in the case of Lagrangian (42) and Hamiltonian (55) it is $\varepsilon^a$ which interacts with matter. This symmetry – a "reflection" symmetry (since $mcu^5$ and $p_5$ may be interchanged) – will be examined in more detail in later sections.

## V.  THE FIELD EQUATIONS

The field equations are

$$\partial_a \frac{\partial(\mathcal{L}\varphi)}{\partial_a \phi_n} = \frac{\partial(\mathcal{L}\varphi)}{\partial \phi_n} \tag{85}$$

for fields $\phi_n$, where $\mathcal{L} = \mathcal{L}_{field} + \mathcal{L}_{matter}$. Care must be used when generating the field equations for $\gamma_{ab}$, $\varepsilon_a$, $\gamma^{ab}$ and $\varepsilon^a$ since these quantities are not independent of each other. Therefore introduce a new term

$$\beta_b^a (\delta_a^b - \gamma^{bc}\gamma_{ca} - \varepsilon^b \varepsilon_a) + \beta^a \gamma_{ab}\varepsilon^b + \beta_a \gamma^{ab}\varepsilon_b \tag{86}$$

to the Lagrangian, where the $\beta_b^a$, $\beta^a$ and $\beta_a$ are Lagrange multipliers, so that $\gamma_{ab}$, $\varepsilon_a$, $\gamma^{ab}$ and $\varepsilon^a$ can be treated as being independent when fields are being varied. The Lagrangian multipliers can then be eliminated from the resulting equations leaving just equations for the field. The raw equations are

$$\partial_a \frac{\partial(\mathcal{L}\varphi)}{\partial_a \gamma_{bc}} = \frac{\partial(\mathcal{L}\varphi)}{\partial \gamma_{bc}} - \tfrac{1}{2}\varphi(\beta_a^b \gamma^{ac} + \beta_a^c \gamma^{ab}) + \tfrac{1}{2}\varphi(\beta^b \varepsilon^c + \beta^c \varepsilon^b) \tag{87}$$

$$\partial_a \frac{\partial(\mathcal{L}\varphi)}{\partial_a \gamma^{bc}} = \frac{\partial(\mathcal{L}\varphi)}{\partial \gamma^{bc}} - \tfrac{1}{2}\varphi(\beta^a_b \gamma_{ac} + \beta^a_c \gamma_{ab}) + \tfrac{1}{2}\varphi(\beta_b \varepsilon_c + \beta_c \varepsilon_b) \tag{88}$$

$$\partial_a \frac{\partial(\mathcal{L}\varphi)}{\partial_a \varepsilon_b} = \frac{\partial(\mathcal{L}\varphi)}{\partial \varepsilon_b} - \varphi \beta^b_a \varepsilon^a + \varphi \beta_a \gamma^{ab} \tag{89}$$

and

$$\partial_a \frac{\partial(\mathcal{L}\varphi)}{\partial_a \varepsilon^b} = \frac{\partial(\mathcal{L}\varphi)}{\partial \varepsilon^b} - \varphi \beta^a_b \varepsilon_a + \varphi \beta^a \gamma_{ab}. \tag{90}$$

Then equation (87) contracted with $h^d_b h^e_c$ minus equation (88) contracted with $\gamma^{bd}\gamma^{ce}$ gives

$$\partial_a\left(\frac{\partial(\mathcal{L}\varphi)}{\partial_a \gamma_{bc}}\right) h^d_b h^e_c - \partial_a\left(\frac{\partial(\mathcal{L}\varphi)}{\partial_a \gamma^{bc}}\right)\gamma^{bd}\gamma^{ce} = \frac{\partial(\mathcal{L}\varphi)}{\partial \gamma_{bc}} h^d_b h^e_c - \frac{\partial(\mathcal{L}\varphi)}{\partial \gamma^{bc}}\gamma^{bd}\gamma^{ce}; \tag{91}$$

equation (87) contracted with $\varepsilon_b h^e_c$ minus $\tfrac{1}{2}$ times equation (90) contracted with $\gamma^{be}$ gives

$$\partial_a\left(\frac{\partial(\mathcal{L}\varphi)}{\partial_a \gamma_{bc}}\right)\varepsilon_b h^e_c - \tfrac{1}{2}\partial_a\left(\frac{\partial(\mathcal{L}\varphi)}{\partial_a \varepsilon^b}\right)\gamma^{be} = \frac{\partial(\mathcal{L}\varphi)}{\partial \gamma_{bc}}\varepsilon_b h^e_c - \tfrac{1}{2}\frac{\partial(\mathcal{L}\varphi)}{\partial \varepsilon^b}\gamma^{be}; \tag{92}$$

equation (88) contracted with $\varepsilon^b h^c_e$ minus $\tfrac{1}{2}$ times equation (89) contracted with $\gamma_{be}$ gives

$$\partial_a\left(\frac{\partial(\mathcal{L}\varphi)}{\partial_a \gamma^{bc}}\right)\varepsilon^b h^c_e - \tfrac{1}{2}\partial_a\left(\frac{\partial(\mathcal{L}\varphi)}{\partial_a \varepsilon_b}\right)\gamma_{be} = \frac{\partial(\mathcal{L}\varphi)}{\partial \gamma^{bc}}\varepsilon^b h^c_e - \tfrac{1}{2}\frac{\partial(\mathcal{L}\varphi)}{\partial \varepsilon_b}\gamma_{be}; \tag{93}$$

while equation (89) contracted with $\varepsilon_b$ minus equation (90) contracted with $\varepsilon^b$ gives

$$\partial_a\left(\frac{\partial(\mathcal{L}\varphi)}{\partial_a \varepsilon_b}\right)\varepsilon_b - \partial_a\left(\frac{\partial(\mathcal{L}\varphi)}{\partial_a \varepsilon^b}\right)\varepsilon^b = \frac{\partial(\mathcal{L}\varphi)}{\partial \varepsilon_b}\varepsilon_b - \frac{\partial(\mathcal{L}\varphi)}{\partial \varepsilon^b}\varepsilon^b. \tag{94}$$

From equation (91) one finds

$$\tfrac{1}{2}(R^a_b h^e_a \gamma^{bf} + R^a_b h^f_a \gamma^{be}) - \omega \Phi_a \Phi_b \gamma^{ae}\gamma^{bf} - \omega \lambda_0 \gamma^{ae}\gamma^{bf}\gamma^{cd} E_{ac} E_{bd}$$
$$- \tfrac{1}{2}\gamma^{ef}(R - 2\Lambda - \omega \gamma^{ab}\Phi_a \Phi_b - \tfrac{1}{2}\omega \lambda_0 \gamma^{ab}\gamma^{cd} E_{ac} E_{bd})$$
$$= \frac{16\pi G_0}{c^3}\sum \frac{1}{2mc} p_a p_b \gamma^{ae}\gamma^{bf}\rho \tag{95}$$

where the sum is over all particles, and from equation (94) one finds

$$R - 2\Lambda + 2(\omega+1)\varphi^{-1}\partial_a(\varphi\gamma^{ab}\Phi_b) - \omega\gamma^{ab}\Phi_a\Phi_b + \tfrac{1}{2}\lambda_0^{-1}\gamma_{ab}\gamma_{cd}E^{ac}E^{bd}$$
$$+ \lambda_0^{-1}\varepsilon^a(\partial_a E) - \tfrac{1}{2}(3\omega+1)\lambda_0\gamma^{ab}\gamma^{cd}E_{ac}E_{bd} = 0. \tag{96}$$

Contracting equation (95) with $\gamma_{ef}$ and combining with equation (96) yields a further scalar equation

$$(2\omega+3)\varphi^{-1}\partial_a(\varphi\gamma^{ab}\Phi_b) + 2\Lambda + \tfrac{3}{4}\lambda_0^{-1}\gamma_{ab}\gamma_{cd}E^{ac}E^{bd} + \tfrac{3}{2}\lambda_0^{-1}\varepsilon^a(\partial_a E)$$
$$-\tfrac{3}{4}(2\omega+1)\lambda_0\gamma^{ab}\gamma^{cd}E_{ac}E_{bd} = -\frac{8\pi G_0}{c^3}\sum mc\rho. \qquad (97)$$

Equation (97) is the equivalent of the equation for the scalar field in Brans-Dicke theory[15]. Deduce that $\omega$ cannot be $-\tfrac{3}{2}$, otherwise – with the exception of the term in $\varepsilon^a(\partial_a E)$ – there would be no second derivatives in equation (97), and so it would represent a constraint when viewed in four dimensions.

Continuing with the field equations, from equation (92) one finds

$$\tfrac{1}{2}(\nabla_a E^{ab})h_b^e - \tfrac{1}{2}(\partial_a E)\gamma^{ae} + \tfrac{1}{2}\varepsilon^a(\nabla_a\Phi_b)\gamma^{be} - \tfrac{1}{2}\lambda_0(\nabla_a E_{bc})\gamma^{ab}\gamma^{ce}$$
$$-\tfrac{1}{2}(2\omega+1)\lambda_0\gamma^{ab}\Phi_a E_{bc}\gamma^{ce} = 0 \qquad (98)$$

assuming the equations of motion are derived from the Hamiltonian (47); in the case where the equations of motion are derived from the Lagrangian (42) the right-hand side is

$$\frac{16\pi G_0}{c^3}\sum \tfrac{1}{2}mc(\varepsilon_a u^a)u^b h_b^e \rho.$$

From equation (93) one finds

$$\tfrac{1}{2}(\nabla_a E^{ab})h_b^e - \tfrac{1}{2}(\partial_a E)\gamma^{ae} + \tfrac{1}{2}(2\omega+1)\varepsilon^a(\nabla_a\Phi_b)\gamma^{be} - \tfrac{1}{2}(2\omega+1)\lambda_0(\nabla_a E_{bc})\gamma^{ab}\gamma^{ce}$$
$$-\tfrac{1}{2}(4\omega+1)\lambda_0\gamma^{ab}\Phi_a E_{bc}\gamma^{ce} = \frac{16\pi G_0}{c^3}\sum\frac{1}{2mc}(\varepsilon^a p_a)p_b\gamma^{be}\rho \qquad (99)$$

assuming the equations of motion are derived from the Hamiltonian (47); in the case where the equations of motion are derived from the Lagrangian (42) the right-hand side is zero. Note that the left-hand side of equation (98) is the same as the left-hand side of equation (99) when $\omega=0$. Since the right-hand sides of these equations are different, it follows that $\omega$ cannot be 0.

Equations (98) and (99) can be resolved into

$$(\nabla_a E^{ab})h_b^e - (\partial_a E)\gamma^{ae} - 2\omega\lambda_0\gamma^{ab}\Phi_a E_{bc}\gamma^{ce}$$
$$= -\frac{8\pi G_0}{\omega c^3}\sum\frac{1}{mc}(\varepsilon^a p_a)p_b\gamma^{be}\rho, \qquad (100)$$

and

$$\lambda_0(\nabla_a E_{bc})\gamma^{ab}\gamma^{ce} - \varepsilon^a(\nabla_a\Phi_b)\gamma^{be} + \lambda_0\gamma^{ab}\Phi_a E_{bc}\gamma^{ce}$$
$$= -\frac{8\pi G_0}{\omega c^3}\sum\frac{1}{mc}(\varepsilon^a p_a)p_b\gamma^{be}\rho. \qquad (101)$$

In the case where the equations of motion are derived from the Lagrangian (42) the right-hand side of equations (100) and (101) are

$$\frac{8\pi G_0(2\omega+1)}{\omega c^3}\sum mc(\varepsilon_a u^a)u^b h_b^e \rho$$

and

$$\frac{8\pi G_0}{\omega c^3} \sum mc(\varepsilon_a u^a) u^b h_b^e \rho$$

respectively. Note the effect of the term $-(\partial_a E)\gamma^{ae}$ in equation (100): in the (4+1)-dimensional notation $E$ is $2(\partial_\mu \varepsilon^\mu)$ so that $(\nabla_a E^{ab})h_b^e - (\partial_a E)\gamma^{ae}$ is $\partial_\mu (g^{\mu\nu}(\partial_\nu \varepsilon^\rho) - g^{\rho\nu}(\partial_\nu \varepsilon^\mu))$ (assuming weak fields, etc.). Therefore, even though $E^{ab}$ is a symmetric tensor, these are still Maxwell's equations (to first order).

The constant $\lambda_0$ may be set to $\pm 1$ by scaling the fifth dimension with a constant, i.e. $w \to w \times$ constant. Assume that $\lambda_0 = 1$ and $\omega > 0$ so that both photons have positive energy. Then in the case where the equations of motion are derived from the Hamiltonian (47), deduce from equation (101) that in order to recover Coulomb's law

$$\kappa = \sqrt{\frac{\omega \mu_0 c^4}{8\pi G_0}} \ , \tag{102}$$

where $\mu_0$ is the permeability constant; in the case where the equations of motion are derived from the Lagrangian (42) deduce that

$$\kappa = \sqrt{\frac{\omega \mu_0 c^4}{8\pi G_0 (2\omega + 1)}} \ . \tag{103}$$

The constant $G_0$ is arranged to recover Newton's law of gravitation. It is not the same as Newton's constant of gravitation because the scalar field acts as a source of gravitation and therefore alters the results of general relativity, as in the Brans-Dicke theory. In the spherically symmetric solution of the Brans-Dicke equations one finds

$$g_{00} = -1 + \frac{2G_0 m}{c^2 r}\left(\frac{2\omega + 4}{2\omega + 3}\right) \tag{104}$$

and so the value of $G_0$ must be set so as to offset the additional factor of $(2\omega+4)/(2\omega+3)$, i.e.

$$G = \frac{2\omega + 4}{2\omega + 3} G_0 . \tag{105}$$

Other effects, such as the deflection of light and the perihelion advance of Mercury, depend on the ratio $g_{00}/g_{ii}$, and these allow a limit to be set on the constant $\omega$: experiment currently puts the value of $\omega$ at greater than 40,000 (Ref. 16). Generally speaking, the Brans-Dicke theory tends towards Einstein's theory in the limit as $\omega \to \infty$ (though there are exceptions to this rule – for example, see Ref. 17); the equivalent statement in the theory of degenerate metrics is that the theory tends towards the Einstein-Maxwell theory in the limit as $\omega \to \infty$. In the case of the Hamiltonian (47) $\varepsilon_a \sim 1/\sqrt{\omega}$ and $\varepsilon^a \sim 1/\sqrt{\omega}$ as $\omega \to \infty$, while in the case of the Lagrangian (42) $\varepsilon_a \sim 1/\omega$ and $\varepsilon^a \sim 1$; in both cases the total electromagnetic energy is independent of $\omega$, as must be the case.

Lastly, the Noether current associated with eigengauge invariance is

$$\lambda_0 \gamma^{ab} \gamma^{cd} E_{bc}(\partial_d \alpha) - \varepsilon^a \gamma^{bc} \Phi_b (\partial_c \alpha) \ ;$$

the divergence of this vector is the left-hand side of equation (101) contracted with $(\partial_e \alpha)$, which is zero in a vacuum.

## VI. THE NATURE OF THE FIFTH DIMENSION

When one thinks of the time dimension from the point of view of matter, one thinks of particles as being spread out uniformly over the whole of time. A particle will visit all points in time – if it is not destroyed first – whereas it cannot be said that it will visit all points in space. This behaviour is encapsulated in the density function $\rho$ which, in the classical world, can be expressed as

$$\rho(x^\mu) = \delta(x - x(ct))\delta(y - y(ct))\delta(z - z(ct)) \, .$$

The density function obeys the continuity equation

$$\nabla_\mu (u^\mu \rho) = 0 \, .$$

The question arises – what is the nature of the fifth dimension? Particles described by the Hamiltonian (47) obey $\varepsilon_a u^a = 0$, i.e. they remain at fixed values of $w$. It is clear that the fifth dimension cannot be like the three dimensions of space, with a particle existing at one point along the dimension at any particular instant, since we would notice that directly (assuming particles are not all coincident at a single value of $w$); it must be like time, with particles spread out uniformly along it. In Kaluza-Klein theory the cylinder condition $\partial_5 \gamma^{ab} = 0$ imposes this property by eliminating dependency on the fifth coordinate, however the existence of a preferred coordinate breaks the unification. The correct relativistic equation would be $E^{ab} = 0$, since $E^{ab}$ contains the term $-\varepsilon^c (\partial_c \gamma^{ab})$, but $E^{ab}$ is needed to recover Coulomb's law, so it cannot be zero. Something else is needed to replace the cylinder condition.

Let us address this problem by considering what happens to the density function $\rho$ in five dimensions. In five dimensions the continuity equation is

$$\nabla_a (u^a \rho) = 0 \, . \tag{106}$$

Now consider what equation (106) looks like for a particle at rest in flat space. For a particle described by the Hamiltonian (47) $u^a = (1, 0, 0, 0, 0)$ and thus

$$\frac{1}{c} \frac{\partial \rho}{\partial t} = 0 \, . \tag{107}$$

Therefore $\rho = \rho(x, y, z, w)$, determined by an initial state at time $t = t_0$. However if matter is to be spread out across the fifth dimension, there must be an additional condition which implies that

$$\frac{\partial \rho}{\partial w} = 0 \, , \tag{108}$$

i.e. in relativistic terms either

$$\nabla_a (\varepsilon^a \rho) = 0 \tag{109}$$

or

$$\varepsilon^a \partial_a \rho = 0 \, . \tag{110}$$

Fortunately, instead of having to impose a condition arbitrarily, it is possible to derive a suitable equation by using the variational principle.

Let us return to the eigengauge transformation and consider what happens when the constraint on $\alpha$ given by equation (67) is removed. The transformation then becomes

$$\gamma_{ab} \to \gamma_{ab} - \lambda_0 h_a^c (\partial_c \alpha) \varepsilon_b - \lambda_0 h_b^c (\partial_c \alpha) \varepsilon_a$$

$$\varepsilon_a \to \lambda^{-1} \varepsilon_a + (\partial_a \alpha)$$

$$\varepsilon^a \to \varepsilon^a + \lambda_0 \gamma^{ac} (\partial_c \alpha) \tag{111}$$

$$\gamma^{ab} \to \gamma^{ab} - \gamma^{ac} (\partial_c \alpha) \varepsilon^b - \gamma^{bc} (\partial_c \alpha) \varepsilon^a$$

where for simplicity $\alpha$ is assumed to be small so that terms in second order and above can be ignored; $\lambda = 1 + \varepsilon^c (\partial_c \alpha)$ as before. Under (111) $\varphi$ is still invariant, but the Lagrangian (75) transforms as

$$\frac{16\pi G_0}{c^3} \mathcal{L} \to \frac{16\pi G_0}{c^3} \mathcal{L} + (2\omega - 1)(\partial_a \lambda) \gamma^{ab} \Phi_b$$
$$+ \varepsilon^e (\partial_e \alpha) \left( 2\omega \gamma^{ab} \Phi_a \Phi_b - \frac{1}{\varphi} \partial_a (\varphi \gamma^{ab} \Phi_b) + \omega \lambda_0 \gamma^{ab} \gamma^{cd} E_{ac} E_{bd} \right)$$

plus a divergence, and thus from the variational principle one finds

$$\frac{1}{\varphi} \partial_e \left( \varphi \varepsilon^e \left( \frac{1}{\varphi} \partial_a (\varphi \gamma^{ab} \Phi_b) - \gamma^{ab} \Phi_a \Phi_b - \tfrac{1}{2} \lambda_0 \gamma^{ab} \gamma^{cd} E_{ac} E_{bd} \right) \right) = 0 \tag{112}$$

assuming equations of motion derived from the Hamiltonian (47). Note how the effect of transforming the Lagrangian in this way results in terms involving $\varepsilon^c (\partial_c \alpha)$ which, when varied, gives an equation which is of the form $\partial_5 (...) = 0$. The quantity inside the bracket in equation (112) is independent of the fifth dimension. This equation, then, replaces the cylinder condition, except now, rather than being a condition, it is simply a consequence of the field equations. Combining equation (112) with equation (97) one obtains

$$\frac{1}{\varphi} \partial_e \left( \varphi \varepsilon^e \left( \tfrac{3}{4} \lambda_0^{-1} \gamma_{ab} \gamma_{cd} E^{ac} E^{bd} + \tfrac{3}{2} \lambda_0^{-1} \varepsilon^a (\partial_a E) - \tfrac{1}{4} (2\omega - 3) \lambda_0 \gamma^{ab} \gamma^{cd} E_{ac} E_{bd} \right. \right.$$
$$\left. \left. + (2\omega + 3) \gamma^{ab} \Phi_a \Phi_b + \frac{8\pi G_0}{c^3} \sum mc\rho \right) \right) = 0. \tag{113}$$

With no second derivatives inside the bracket other than $\varepsilon^a (\partial_a E)$ the implication is that $\nabla_a (\varepsilon^a \rho) = 0$ and therefore matter is spread out uniformly across the fifth dimension, i.e. the fifth dimension is not observable under ordinary conditions, consistent with experience. In the case where the equations of motion are derived from the Lagrangian (42) an additional term in $q^2/m$ arises since

$$L \to L - \lambda_0 mc (\varepsilon_a u^a)(\partial_b \alpha) u^b + \lambda_0 mc (\varepsilon_a u^a)^2 (\partial_b \alpha) \varepsilon^b$$

under (111), however the impact of this term may be eliminated by modifying the continuity equation to be $\nabla_a (h_b^a u^b \rho) = 0$, so that equation (112) then still holds. In both cases – Hamiltonian

(47) and Lagrangian (42) – the particle is a sheet in five dimensions, extending across time $t$ and the fifth dimension $w$. Thus a particle is no longer a world-line, it is a "world-sheet".

Let us now return to equation (36): $E^{ab}p_a p_b = 0$. Clearly it cannot be the case that simply $E^{ab} = 0$ since that would imply no charges – $\varepsilon^a$ has a source whichever equations of motion are used. Therefore the field equations must allow some non-trivial solutions for $E^{ab}$ which nevertheless satisfy equation (36) at any point in space-time where there is a particle.

The key to understanding how equation (36) is solved is to consider what happens to $E^{ab}p_a p_b$ under the eigengauge transformation. Then

$$E^{ab}p_a p_b \to E^{ab}p_a p_b + 2\lambda_0(\nabla_a \partial_b \alpha)\gamma^{ac}\gamma^{bd}p_c p_d$$
$$+ 2\lambda_0 E_{ab}\gamma^{ac}\gamma^{bd}(\partial_c \alpha)p_d(\varepsilon^e p_e) - 2E^{ab}(\partial_a \alpha)p_b(\varepsilon^c p_c) + O(\alpha^2). \quad (114)$$

(In order for equation (114) to have meaning, the 5-momentum $p_a$ must be extended in some continuous way beyond the world-sheet of the particle.) The presence of the second derivative of $\alpha$ in equation (114) means there is no difficulty in solving $E^{ab}p_a p_b = 0$, since this equation can simply be viewed as an equation for $\alpha$ with the boundary condition that $\partial_a \alpha = 0$ on the particle world-sheet (since the equations of motion are not eigengauge invariant). Note how equation (36), which was derived by varying the particle, is solved by the field; while (for example) equation (81), which, in the case of the Hamiltonian (47), was derived by varying the field by the eigengauge transformation, is solved by the particle.

In section 3 it was shown how Maxwell's equations are recovered from the equations of motion, but only for weak fields. Now that it has been shown that a particle is a world-sheet, it is possible to improve the treatment, by showing that Maxwell's equations are in fact recovered exactly – as one would expect, given the accuracy to which Maxwell's theory has been verified.

As before, work in the (4+1)-dimensional notation with $g^{\mu\nu} = \eta^{\mu\nu}$ but now without the requirement for the electromagnetic fields to be weak. Whichever Lagrangian or Hamiltonian is used, the particle will obey $\gamma_{ab}u^a u^b = -1$, and so expanding this out

$$\eta_{\mu\nu}u^\mu u^\nu - 2\eta_{\mu\rho}\varepsilon^\rho u^\mu(\varepsilon_\nu u^\nu + \varepsilon_5 u^5) + (\eta_{\rho\sigma}\varepsilon^\rho \varepsilon^\sigma)(\varepsilon_\mu u^\mu + \varepsilon_5 u^5)^2 = -1. \quad (115)$$

In the case where the equations of motion are derived from the Lagrangian (42) the particle would appear to be "off mass-shell" in four dimensions, since $\varepsilon_a u^a$ is non-zero. However if $\varepsilon^\mu = 0$ on the particle world-sheet, then the particle becomes "on mass-shell", i.e. $\eta_{\mu\nu}u^\mu u^\nu = -1$. Then one finds that

$$\eta_{\mu\nu}\frac{du^\nu}{ds} = -(\partial_\mu(g_{\nu\rho}\varepsilon^\rho) - \partial_\nu(g_{\mu\rho}\varepsilon^\rho))u^\nu(\varepsilon_\sigma u^\sigma + \varepsilon_5 u^5)$$
$$+ \eta_{\mu\rho}(\partial_5 \varepsilon^\rho)u^5(\varepsilon_\sigma u^\sigma + \varepsilon_5 u^5). \quad (116)$$

The term in $(\partial_5 \varepsilon^\mu)$ may be seen to be zero because the particle world-sheet obeys $\partial_5 \rho = 0$ and $\varepsilon^\mu$ is constant on this surface. Thus Maxwell's equations are in fact recovered exactly – the weak-field approximation is not needed. In the case where the equations of motion are derived from the Lagrangian (28) the particle is automatically on mass-shell, since $\varepsilon_a u^a = 0$. One finds that

$$\eta_{\mu\nu}\frac{du^\nu}{ds} = \frac{\kappa q}{mc}((\partial_\mu \varepsilon_\nu) - (\partial_\nu \varepsilon_\mu))u^\nu + \frac{\kappa q}{mc}((\partial_\mu \varepsilon_5) - (\partial_5 \varepsilon_\mu))u^5 + \eta_{\nu\rho}\varepsilon^\nu u^\rho(\partial_5 \varepsilon_\mu)u^5$$
$$- \eta_{\rho\sigma}\varepsilon^\rho u^\sigma((\partial_\mu \varepsilon_\nu) - (\partial_\nu \varepsilon_\mu))u^\nu + \varepsilon_\mu \eta_{\nu\sigma}(\partial_5 \varepsilon^\sigma)u^\nu u^5 + \varepsilon_\mu \eta_{\nu\sigma}(\partial_\rho \varepsilon^\sigma)u^\nu u^\rho. \quad (117)$$

As before, set $\varepsilon^\mu = 0$ on the particle world-sheet to simplify to

$$\eta_{\mu\nu}\frac{du^{\nu}}{ds}=\frac{\kappa q}{mc}((\partial_{\mu}\varepsilon_{\nu})-(\partial_{\nu}\varepsilon_{\mu}))u^{\nu}+\frac{\kappa q}{mc}((\partial_{\mu}\varepsilon_{5})-(\partial_{5}\varepsilon_{\mu}))u^{5}. \tag{118}$$

Again Maxwell's equations are recovered exactly, however there is now a new term in $(\partial_{\mu}\varepsilon_{5})-(\partial_{5}\varepsilon_{\mu})$ which is the field strength $\Phi_{\mu}$. Thus the reflection symmetry is not exact, however since $\Phi_{\mu}$ is a gravitational field strength, its impact will be negligible. Relative to electromagnetic effects, the term in $\Phi_{\mu}$ will be of the order of $g_0 R_E c^{-2}\omega^{-1}$ at the surface of the Earth, where $g_0$ is the acceleration due to gravity and $R_E$ is the radius of the Earth. The effect is therefore of the order of $10^{-10}/\omega$. (Perhaps one can argue that, in order for the reflection symmetry to be exact, $\omega$ must tend to infinity.) Similarly there is some impact from $\Phi_{\mu}$ on the two vector field equations (100) and (101), but it is no greater than the usual impact from gravitation on Maxwell's equations.

As for setting $\varepsilon^{\mu}=0$ on the particle world-sheet, speculate that this may be done with a suitable coordinate transformation of the type $x^{\mu}\to x^{\mu'}(x^{\mu},w)$. The situation is analogous to the surface of a sphere in three-dimensional space. One may transform from Cartesian coordinates ($x$, $y$, $z$) to spherical polar coordinates when the "density" function of the surface becomes $\rho=\delta(r-r_0)$ and satisfies $\partial\rho/\partial\theta=0$ and $\partial\rho/\partial\phi=0$. If one thinks of $\theta$ as $t$ and $\phi$ as $w$, then $\partial\rho/\partial w=0$.

Regarding the observational consequences of the second photon, there is no knowing how much, or how little dark light there may be in the universe, since by its very nature it does not interact with matter. There is no cosmological measure against which the total energy of dark light might be scaled – and there is no reason to suppose that the sum of dark light is the same as the sum of radiant light. Therefore the existence of the second photon does not contradict any known astrophysical observations.

## VII. MASSIVE PHOTONS AND DARK MATTER

Since equation (112) is a scalar equation, and therefore less restrictive than the cylinder condition (which is a matrix expression), it is natural to ask whether there are any phenomena that can be explained by making use of the dependency of the field on the fifth dimension.

In this section work in (4+1)-dimensional notation, and for simplicity assume that there is no gravitational field, i.e. $g^{\mu\nu}=\eta^{\mu\nu}$, and $\varepsilon_5=1$ so that $\varphi=1$. Then in a vacuum equation (101) becomes

$$\eta^{\mu\nu}\partial_{\mu}(\partial_{\nu}\varepsilon_{\rho}-\partial_{\rho}\varepsilon_{\nu})+\partial_5(\partial_5\varepsilon_{\rho})=0 \tag{119}$$

while equation (100) becomes

$$\eta^{\mu\nu}(\partial_{\mu}(\partial_{\nu}\eta_{\rho\sigma}\varepsilon^{\sigma})-\partial_{\mu}(\partial_{\rho}\eta_{\nu\sigma}\varepsilon^{\sigma}))+\partial_5(\partial_5\varepsilon_{\rho})=0 \tag{120}$$

assuming that $\lambda_0=1$. The terms that give Maxwell's equations are evident in both equations (119) and (120).

Equation (120) is unusual because there is no term in $\partial_5(\partial_5\varepsilon^{\rho})$. A further equation may be obtained from equation (91):

$$\partial_5(\partial_{\mu}\varepsilon_{\nu})+\partial_5(\partial_{\nu}\varepsilon_{\mu})=\partial_5(\partial_{\mu}\eta_{\nu\sigma}\varepsilon^{\sigma})+\partial_5(\partial_{\nu}\eta_{\mu\sigma}\varepsilon^{\sigma}), \tag{121}$$

but again there is no second derivative of $\varepsilon^{\mu}$ in the fifth dimension. Thus $\varepsilon^{\mu}$ is only determined up to an arbitrary function of $w$ which can be added to it. However, since it was argued in section 6 that the terms in $(\partial_5\varepsilon^{\mu})$ in the equations of motion can be transformed away, this arbitrary function of $w$ is of no importance.

There are two distinct solutions to equation (119). The first solution is independent of the fifth coordinate $w$ and is the familiar photon

$$\varepsilon_\mu = a_\mu \exp(ik_\nu x^\nu) \qquad (122)$$

where $\eta^{\mu\nu} a_\mu k_\nu = 0$ and $\eta^{\mu\nu} k_\mu k_\nu = 0$. The second solution allows for the wave behaviour to extend to the fifth dimension

$$\varepsilon_\mu = a_\mu \exp(ik_\nu x^\nu + ik_5 x^5) \qquad (123)$$

where $\eta^{\mu\nu} a_\mu k_\nu = 0$ and now $\eta^{\mu\nu} k_\mu k_\nu + (k_5)^2 = 0$. Thus this solution represents a massive photon: the mass is determined by the frequency of the vibration in the fifth dimension. Note that there is a continuous spectrum of possible photon masses – the photon mass is not a discrete value like the mass of the electron or the mass of the proton. Since $k_\mu$ is a timelike vector and $\eta^{\mu\nu} a_\mu k_\nu = 0$ it follows that $a_\mu$ must be a spacelike vector.

There are also two distinct solutions to equation (120). The first solution is the familiar photon

$$\varepsilon^\mu = \tilde{a}^\mu \exp(i\tilde{k}_\nu x^\nu) \qquad (124)$$

where $\tilde{a}^\mu \tilde{k}_\mu = 0$ and $\eta^{\mu\nu} \tilde{k}_\mu \tilde{k}_\nu = 0$, provided that $\partial_5 \varepsilon_\rho = 0$. If $\partial_5 \varepsilon_\rho$ is non-zero, there is a second solution

$$\varepsilon^\mu = \eta^{\mu\rho} a_\rho \exp(ik_\nu x^\nu + ik_5 x^5) \qquad (125)$$

which is the same massive photon as $\varepsilon_\mu$. In either case – whether $\partial_5 \varepsilon_\mu$ is zero or non-zero – the reflection symmetry is seen to hold. Note that the one massive photon creates the other – they must both exist at the same time. In general, if $\varepsilon_\rho = \varepsilon_\rho(x^\mu)\exp(ik_5 x^5)$ then equation (119) becomes the Proca equation:

$$\eta^{\mu\nu} \partial_\mu (\partial_\nu \varepsilon_\rho - \partial_\rho \varepsilon_\nu) = (k_5)^2 \varepsilon_\rho, \qquad (126)$$

and similarly for equation (120). Note, however, that while the Proca equation in four dimensions violates gauge invariance, there is no such violation in five dimensions, since the mass term $\partial_5(\partial_5 \varepsilon_\rho)$ in equations (119) and (120) is invariant under $\varepsilon_\rho \to \varepsilon_\rho - \partial_\rho f$ where $f = f(x^\mu)$.

Regarding the interaction of massive photons with matter, the effect of the spreading out of matter across the fifth dimension is to average across the fifth dimension. Because massive photons are waves in the fifth dimension, the result of the averaging is zero, so that there is no interaction between massive photons and normal matter. Massive photons are "dark".

However, just because massive photons are dark does not mean that they have no visible effects. Consider the energy-momentum tensor of the field, which can be obtained from the Lagrangian. Evaluating the various scalar quantities for the massive photons, we have

$$R = 0; \qquad (127)$$

$$\varphi^{-1} \partial_a (\varphi \gamma^{ab} \Phi_b) = 0; \qquad (128)$$

$$\gamma^{ab} \gamma^{cd} E_{ac} E_{bd} = 2(\eta^{\nu\rho} k_\nu k_\rho)(\eta^{\sigma\tau} a_\sigma a_\tau) \cos^2(k_\mu x^\mu + k_5 x^5); \qquad (129)$$

$$\gamma^{ab} \Phi_a \Phi_b = (k_5)^2 (\eta^{\sigma\tau} a_\sigma a_\tau) \cos^2(k_\mu x^\mu + k_5 x^5); \qquad (130)$$

$$\gamma_{ab}\gamma_{cd}E^{ac}E^{bd} = 2(\eta^{\nu\rho}k_\nu k_\rho)(\eta^{\sigma\tau}a_\sigma a_\tau)\cos^2(k_\mu x^\mu + k_5 x^5); \tag{131}$$

and

$$E^2 = 0. \tag{132}$$

Note that the complex wave function has been replaced by a real sine function since the wave has been squared. [As a consistency check, one may consider how equation (112) is solved for the massive photon. Since the massive photon is a wave in the fifth dimension, but the quantity inside the bracket in equation (112) is independent of the fifth dimension, the wave-like terms must sum to zero inside the bracket. Substituting in equations (128), (129) and (130), equation (112) may be seen to be satisfied using the relation $\eta^{\mu\nu}k_\mu k_\nu + (k_5)^2 = 0$.] Using equations (127)-(132) the (four-dimensional) energy-momentum tensor of the massive photon is found to be

$$T_{\nu\rho} = \frac{\omega c^3}{16\pi G_0} k_\nu k_\rho (\eta^{\sigma\tau}a_\sigma a_\tau)\cos^2(k_\mu x^\mu + k_5 x^5). \tag{133}$$

Since $T_{00}$ is positive, the massive photon acts as a source of gravity with positive mass. The averaging of the wave over the fifth dimension will introduce a factor of one half. Conclude that the massive photon may provide an explanation for the effects of dark matter.

### VIII. CONCLUSION

It has been shown in this paper how gravitation and electromagnetism can be unified in five dimensions with a degenerate metric, using a gauge-like transformation and a reflection symmetry. The theory makes two predictions: the existence of a second photon which does not interact with matter, and the existence of massive photons which also do not interact directly with matter, but which are candidates for dark matter since they act as a source of gravitation. It is to be hoped that experimental investigations may in time shed some (radiant) light on the validity of the theory.